\documentclass{mn2e}

\usepackage{epsfig} 

\def\gs{\mathrel{\raise0.35ex\hbox{$\scriptstyle >$}\kern-0.6em
\lower0.40ex\hbox{{$\scriptstyle \sim$}}}}
\def\ls{\mathrel{\raise0.35ex\hbox{$\scriptstyle <$}\kern-0.6em
\lower0.40ex\hbox{{$\scriptstyle \sim$}}}}

\begin{document}

\title
[ALMA and magnification bias] 
{Exploiting magnification bias in ultradeep submillimetre-wave surveys 
using ALMA}   
\author
[A.\,W. Blain]
{
A.\,W. Blain
\vspace*{1mm}\\
Institute of Astronomy, Madingley Road, Cambridge, CB3 0HA.\\
Astronomy Department, California Institute of Technology, Pasadena, CA91125, 
USA.\\
}
\maketitle

\begin{abstract}
The surface density of populations of galaxies with steep/shallow source 
counts is increased/decreased by gravitational lensing magnification. These 
effects are usually called `magnification bias' and `depletion' respectively. 
However, if sources are demagnified by lensing, then the situation is 
reversed, and the detectable surface density of galaxies with a shallow 
source count, as expected at the faintest flux densities, is increased. In 
general, demagnified sources are difficult to detect and study: exquisite 
subarcsecond angular resolution and surface brightness sensitivity are 
required, and 
emission from the lensing object must not dominate the image. These 
unusual conditions are expected to be satisfied for observations of the 
dense swarm of demagnified images that could form very close to the line of 
sight 
through the centre of a rich cluster of galaxies using the forthcoming 
submillimetre-wave Atacama Large Millimeter Array (ALMA) interferometer. 
The demagnified images of most of the background galaxies lying within 
about 1\,arcmin of a rich cluster of galaxies could be detected in a single 
18-arcsec-diameter ALMA field centred on the cluster core, providing 
an effective increase in the ALMA field of view. 
This technique could 
allow a representative sample of faint, 10--100\,$\mu$Jy 
submillimetre galaxies to be detected several times more rapidly 
than in a blank field. 
\end{abstract}  

\begin{keywords}
methods: observational -- galaxies: clusters: general -- 
cosmology: observations -- gravitational lensing -- infrared: galaxies --
radio continuum: galaxies
\end{keywords}

\section{Introduction}

Gravitational lensing magnification can have a significant effect on the 
observability of a population of galaxies, via the effect of magnification 
bias. Magnified sources that would otherwise be too faint for detection in 
a practical time can be found (Smail, Ivison \& Blain 1997; 
Altieri et al.\ 1999; 
Pettini et al.\ 2000; Ellis et al.\ 2001), 
and otherwise unresolvable substructure within a source 
can be revealed (Franx et al.\ 1997). Here the effect of magnification bias in 
the innermost core regions of rich clusters of galaxies 
(Broadhurst, Taylor \& Peacock 1995) is 
discussed, in the context of deep observations at very high angular 
resolution using the (sub)millimetre-wave ALMA interferometer 
(Blain 1997, 2001; Wootten 2001)\footnote{Extensive information about 
ALMA can be found at the website http://www.alma.nrao.edu}.
ALMA will be extremely sensitive, but has a small field of view as compared 
with optical and radio telescopes, and so large-area ALMA surveys are 
relatively challenging (Blain 2001). The radius of the ALMA 
field of view is set by the diffraction limit of a 12-m antenna. The 
full-width-half-maximum (FWHM) diameter of the telescope beam ranges from 
about 8\,arcsec at 850\,GHz/350\,$\umu$m to about 1.2\,arcmin at 
90\,GHz/3.3\,mm. 

Here, the de-magnification of lensed images of background galaxies in the 
core of a rich cluster of galaxies is discussed as a tool to enhance the 
efficiency of ALMA to probe the population of very faint submillimetre-wave 
galaxies, as compared with observations in a blank field. 

\section{Magnification bias and depletion} 

When planning a survey, it is important to know how quickly a certain 
number of galaxies can be detected using a telescope. If the galaxies being 
studied are described by a differential source count, in which the surface 
density of galaxies that have intrinsic flux densities between $S$ and 
$S+{\rm d}S$ is $N(S)$, then imposing a gravitational lensing magnification 
factor $\mu$ modifies the count to $N'(S) = N(S/\mu) / \mu^2$. In general, 
$\mu$ is a function of both the redshift and relative position on the sky of 
source and lens. If $N(S)$ can be described by a power law, 
$N(S) \propto S^\alpha$, then the bias factor $B = N'/N = \mu^{-(2+\alpha)}$ 
(Canizares 1981; Borgeest, von Linde \& Refsdal 1991; Schneider 1992). 
$B$ takes a value greater than unity if the magnification bias is positive, 
and a value less than unity if the magnification bias is negative. If sources 
are magnified, that is $\mu > 1$, then the source count is increased if 
$\alpha < -2$, but reduced if $\alpha > -2$. If sources are demagnified, that 
is $\mu < 1$, then these conditions on $\alpha$ are reversed, and so 
a value of $\alpha > -2$ corresponds to a positive magnification bias. 

In almost all studies of high-redshift galaxy populations in which 
gravitational lensing is exploited, magnification rather than demagnification 
is utilized. The single existing 
exception is the use of the relative depletion of red 
galaxies, as compared with blue galaxies, behind rich clusters of galaxies to 
study the cluster potential in the absence of spectroscopic redshifts for all 
of the lensed background galaxies (Broadhurst et al.\ 1995;
Gray et al. 2000; Dye et al.\ 2001). This differential depletion effect arises 
because of the different slopes of the faint counts from band to band: 
compare the slopes of the faint $B$- and $I$-band counts shown in 
Fig.\,1. 

\section{The detection rate of galaxies} 

The importance of magnification bias for a galaxy survey depends on 
several factors. 

First, there is a dependence on the slope of the source counts $\alpha$  
discussed above. The slope of the counts also determines the survey 
strategy that maximizes the detection rate of galaxies. In a fixed observing 
time, it is possible to trade off area coverage and survey depth. Unless it  
is necessary to reach a certain depth in order to detect a specific class 
of objects, this trade off favours a deep 
survey if $\alpha < -3$, and a wide survey if $\alpha > -3$. The appropriate 
trade off in the submillimetre waveband, where count slopes can be very 
steep and change rapidly (Fig.\,1), was discussed by Blain \& Longair (1996). 

Our present knowledge of submillimetre galaxy population (Smail et al.\ 
2001) is that the slope of the 850-$\umu$m counts is close to $\alpha = -3$ 
for flux densities between about 1 and 10\,mJy (Fig.\,1). This indicates that 
existing (sub)millimetre-wave galaxy surveys (Smail et al.\ 1997, 2001; 
Bertoldi et al.\ 2000; Scott et al.\ 2002) 
have been made at the most efficient depth: the detection rate 
is likely to be lower in both deeper and shallower surveys. It is likely, but 
not yet confirmed by observations, that the counts steepen at brighter flux 
densities. This could lead to a very large magnification bias at bright 
($> 100$\,mJy) 850-$\umu$m flux densities (Blain 1997). The counts must 
become shallower, with $\alpha > -2$, at the faintest flux densities;  
otherwise, the sum of the flux density contributed by discrete sources 
would exceed the background radiation intensity measured by  
{\it COBE}-FIRAS (Fixsen et al.\ 1998). 

\begin{figure*}
\begin{center}
\vspace{0.5cm} 
\epsfig{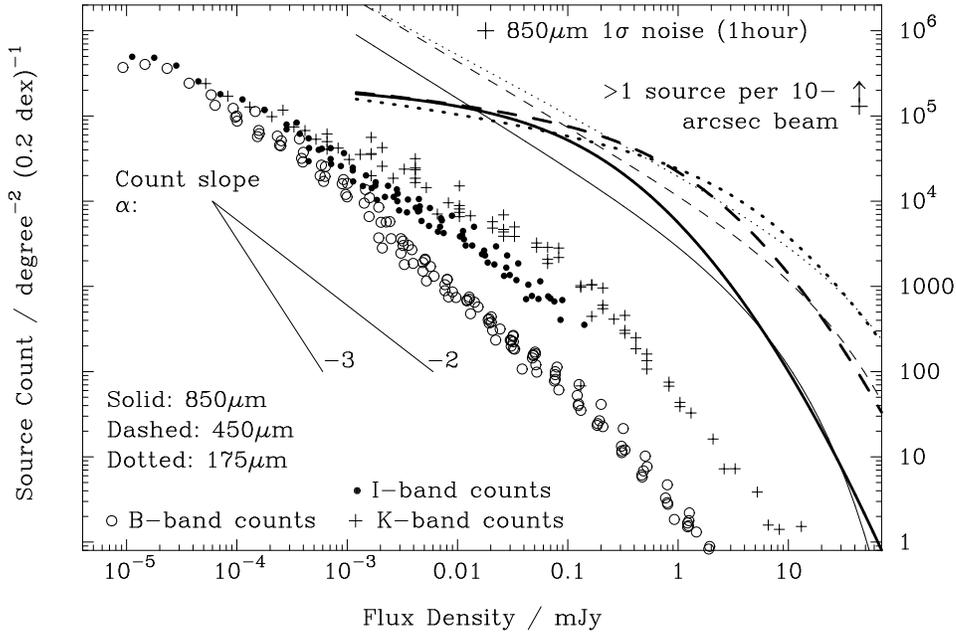}
\end{center}
\caption{Differential source counts in the optical, near- and far-infrared and 
submillimetre wavebands. The $K$-, $I$- and $B$-band data come from the 
compilation of Metcalfe et al.\ (1996) and Maihara et al.\ (2001). Note that 
the faint slope of the K- and B-band counts are considerably different.  
The lines are associated with models for far-infrared and 
submillimetre-wave counts (thick: Blain et al.\ 1999b; thin: Blain et al.\ 
1999c). These models agree with current observational results from SCUBA 
(Smail et al.\ 2001) at 
850\,$\umu$m (from 1 to 10\,mJy) and 450\,$\mu$m (from 10 to 
20\,mJy), and from {\it ISO} at 175\,$\umu$m from 
180 to 500\,mJy (Dole et al.\ 2001). The existing observational constraints are 
imposed at much brighter flux densities than those that ALMA will probe. 
The slope of the 850-$\umu$m counts fainter than about 1\,mJy, and thus 
the magnification bias expected, is currently poorly defined, and awaits the 
results of SMA and ALMA observations. In the most pessimistic case, the 
slope will be similar to that of the faintest optical counts, and so the 
magnification bias will be small. In other models the slope of the faint 
submillimetre-wave counts could be very shallow leading to a strong bias. 
} 
\end{figure*} 

Secondly, the ratio between the instantaneous field of view of a telescope 
and the size of the magnified field affects the significance of magnification 
bias. If the field of view of the telescope is very much larger than the 
magnified region -- as is the case in the optical, radio, and X-ray wavebands, 
and soon in the mid-infrared waveband 
with the launch of {\it SIRTF} -- then magnification 
bias is unlikely to provide a significant assistance to a survey. For example, 
even the optical WFPC-II camera on {\it HST}, with a 2-arcmin field 
of view that is small by current standards, can image 
almost all of the critical 
lensing region of a typical cluster of galaxies in a single pointing (for 
example Smith et al.\ 2001). A single WFPC-II image of a cluster of galaxies 
can be used to probe simultaneously the low-magnification 
($\mu \simeq 1$) regions well outside critical lines, the high-magnification 
($\mu \gg 1$) regions close to the critical lines, and the demagnified 
($\mu < 1$) region well within the critical lines close to the core of the 
cluster. This is even more true for the 3\,arcmin $\times$ 3\,arcmin field of 
view of the forthcoming {\it HST}-ACS camera. 
However, if the field of view is small as compared with the strongly 
magnified area, then even a relatively modest magnification bias can have a 
significant effect. This is especially important if a telescope is only 
sufficiently sensitive to detect a handful of sources in a reasonable 
integration time, as is the case for existing submillimetre-wave 
observations (Smail et al.\ 2001). 

Thirdly, the limit imposed to the maximum depth of a survey due to 
confusion noise can be significant. If the unmagnified population of 
galaxies is too faint to detect above this limit, then the exploitation of 
magnification bias is essential in order to make reliable detections. This 
is the case for the deepest existing submillimetre-wave surveys (Blain, Ivison 
\& Smail 1998). 

\section{Magnification bias and ALMA} 

Although the most efficient detection rate of 850-$\umu$m galaxies is 
likely to be at a depth of 5 to 10\,mJy, it is essential to obtain 
fainter counts, both to probe the properties of sub-$L^*$ high-redshift 
galaxies, and to be sure of the relationship between the counts and the 
integrated intensity of background radiation. It is clear from Fig.\,1 that 
models which provide an adequate description of existing data at mJy 
flux densities (Blain et al.\ 1999b,c) make quite different predictions for 
fainter counts, and so a measurement of very deep submillimetre-wave 
counts could reveal important new information about the evolution of 
high-redshift galaxies. 

Because of source confusion, only interferometers with subarcsec 
resolution, that is ALMA and the Submillimeter Array (SMA; 
Ho 2000)\footnote{Information about SMA can be found at the website   
http://sma2.harvard.edu}, can make these observations. The importance of 
excellent resolution can be seen from the counts in Fig.\,1. At an extreme 
depth of 1\,$\umu$Jy, the surface density of galaxies in the model which 
predicts the greatest count corresponds to only 1 source per 30 0.1-arcsec 
beams. This is a standard definition for a confused image, and a resolution 
limit of 0.1\,arcsec is well within the capabilities of ALMA. 

In a 1-hour integration, within the 18-arcsec-diameter FWHM primary beam, the 
1-$\sigma$ sensitivity of ALMA is 18\,$\umu$Jy at 345\,GHz/870\,$\umu$m 
(Wootten 2001). In a 100-hr integration in a single field, about 20 detections 
would be expected at flux densities brighter than a 5-$\sigma$ threshold 
of 9\,$\umu$Jy. In the same area, a single detection would be expected at 
a 5-$\sigma$ threshold of 0.2\,mJy, corresponding to a 0.2-hr integration. 
Hence, many more galaxies, about 500, could be detected if ALMA were instead 
to map 100 
different fields for 1\,hr each. From a comparison of these results, it is clear 
that a shallower, wider ALMA survey is expected to be more efficient at 
discovering faint submillimetre-wave galaxies. Because of the sensitivity 
of ALMA to CO line emission at very high redshifts (Blain et al.\ 2000), 
an ultradeep pencil-beam redshift survey would be a direct 
by-product. 

Can gravitational lensing be exploited to assist ALMA to probe faint 
submillimetre-wave counts more rapidly? One route would be to exploit the 
high magnifications along critical lines in the image plane of a rich cluster 
of galaxies (Blain 2001) in order to detect the magnified images of very faint 
background galaxies. These would be intrinsically interesting sources, 
regardless of whether magnification bias increases or decreases their 
detection rate. The length of critical lines for a rich cluster at a moderate 
redshift is of order 5\,arcmin, and so about 20 pointings with ALMA at 
345\,GHz would be required to map them. A similarly motivated approach 
would be to image moderate-redshift field galaxies in single deep ALMA 
pointings, especially those classes of galaxies with significant lensing 
cross sections, like massive ellipticals and edge-on disk galaxies (see 
Fig.\,7 in Blain, M\"oller \& Maller 1999), in order to detect strongly-lensed 
magnified images of faint background galaxies. 
Alternatively, it would be possible to exploit the very high angular resolution 
of ALMA to image the densely packed, demagnified counterimages of 
background galaxies that are expected to lie very close to the core of a 
cluster, well within the extent of the critical line structure, and also within 
the diameter of the ALMA primary beam. If the slope of the count of very 
faint background galaxies is flat, with $\alpha > -2$, then the bias factor 
$B$ will be greater than unity. 

The formal description of the lensing properties of the innermost regions of 
a cluster is relatively straightforward. Making the assumption of cylindrical 
symmetry, which is likely to be reasonable, the deflection angle of light 
$\theta_\alpha$ at an impact parameter $r$ depends on the mass enclosed 
$M(<r)$ as $\theta_\alpha \propto M(<r)/r$. If a spherical density profile with 
an index $\xi$, $\rho(r) \propto r^\xi.$, is assumed, then 
$\theta_\alpha \propto r^{\xi+2}$. It is reasonable to assume a constant 
value of the index $\xi$, as we are concerned with only the very central 
regions of clusters. Using the lens equation to relate the 
angular diameter distances connecting the observer, lens and source, the 
magnification  
\begin{equation} 
\mu = \left\vert 1 - {D_{\rm LS} \over D_{\rm OS}} 
\theta_\alpha (\theta_{\rm I})\right\vert^{-1}
\left\vert 1 - (\xi+2){ {D_{\rm LS}} \over {D_{\rm OS}} }
\theta_\alpha (\theta_{\rm I})\right\vert^{-1}, 
\end{equation} 
where $\theta_{\rm I}$ is the angular position of the image. This can be 
re-expressed more simply in terms of the Einstein radius $\theta_{\rm E}$, 
as 
\begin{equation} 
\mu = \left\vert 1 - 
\left({ {\theta_{\rm I}} \over {\theta_{\rm E}} }\right)^{\xi+1} \right\vert
^{-1} \left\vert 1 - (\xi+2) 
\left({ {\theta_{\rm I}} \over {\theta_{\rm E}} }\right)^{\xi+1} 
\right\vert^{-1}
\end{equation}
(Schneider, Ehlers \& Falco 1992).  
The first and second terms yield the conditions for the formation of 
transverse and radial giant arc images respectively. The simplest form of 
the equation occurs for a singular isothermal sphere (SIS) 
with $\xi = -2$, in which 
case the second term vanishes; this is likely to be an extreme lower bound 
on the value of $\xi$. 
Note that the description breaks down if $\xi = -1$, which corresponds 
to the index for a Navarro--Frenk--White (NFW) 
density profile (Navarro, Frenk 
\& White 1997), as derived from halo profiles extracted from 
N-body simulations. In this case,
high-magnification radial-arc images are expected to dominate throughout 
the core of a cluster and almost no demagnified region is expected. 
The presence of dark and baryonic matter associated with a cD 
galaxy in the core of the cluster is sure to generate an index steeper than  
$\xi = -1$ in realistic cases, even if the dark-matter profile is 
described by an NFW profile. Alternative N-body simulations have indicated 
values of $\xi \simeq -1.4$ (Moore et al.\ 1998), while observations 
of X-ray gas profiles (for example Makino \& Asano 1999) and 
{\it HST} images (for example Hammer et al.\ 1997) have been used to 
derive values of $\xi \simeq -1.4$ to $-1.7$ in the central regions 
of clusters. 
The magnification expected as a function of distance from the core of a 
circularly symmetric cluster is compared as a function of $\xi$ in 
Fig.\,2. An SIS produces the most significant demagnification.

\begin{figure}
\begin{center}
\vspace{0.5cm}
\epsfig{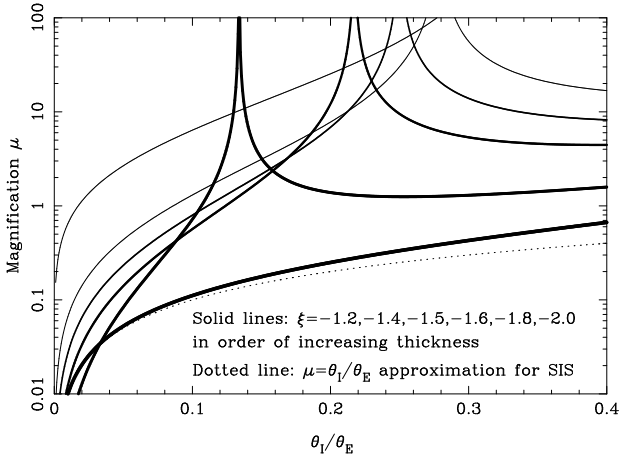}
\end{center}
\caption{The magnification distribution (equation 2) expected as a function 
of angular radius $\theta_{\rm I}$, in units of the Einstein radius 
$\theta_{\rm E}$ for clusters with a range of different central density 
profile indices $\xi$; note that $\xi = -1$ and $-2$
for an NFW profile and SIS respectively. The high-magnification spikes are
due to the formation of radial-arc images, which occur for $\xi > -2$.
Within the radial-arc radius significant demagnifications ($\mu < 1$) 
are expected.}
\end{figure} 

In individual clusters, the magnification 
distribution is certain to be more complex, due to both the gravitational 
potential of the cluster member galaxies and the true aspherical, 
non-isothermal nature of the cluster dark-matter halo; however, in 
reasonable cases, with $\xi \sim -1.5$, strong de-magnification is 
always expected within a 
few arcseconds of the core.  

In order to make a coarse estimate of the maximum size of the effect, it is 
reasonable to assume an SIS radial density profile and  
$\theta_{\rm E} \simeq 40$\,arcsec for a moderate-redshift rich cluster 
similar to Abell 2218. At small radii, 
$\theta_{\rm I} \ll \theta_{\rm E}$, 
$\mu \simeq \theta_{\rm I}/\theta_{\rm E}$ -- see equation 2 and 
the dotted line in 
Fig.\,2 -- and so, as a function
of radius $\theta$, the bias factor 
$B = \mu^{-(2+\alpha)} = (\theta / \theta_{\rm E})^{-(2+\alpha)}$. 
When averaged over a top-hat beam of diameter $\theta_{\rm b}$, 
$\bar B \simeq 2^{3+\alpha}
(\theta_{\rm E}/\theta_{\rm b})^{2+\alpha}/(-\alpha)$, 
while for a Gaussian beam with FWHM diameter $\theta_{\rm b}$, 
$\bar B = (\theta_{\rm E}/\theta_{\rm b})^{2+\alpha} 
(4\, {\rm ln}2)^{1+(\alpha/2)} \Gamma(-\alpha/2)$. 

Values of the effective 
bias $\bar B$ calculated exactly using equation 2 for three different count 
slopes, $\alpha=-1.84$ (Blain et al.\ 
1999c), -1.52 (Blain et al.\ 1999b) and -1.3, and for three different FWHM 
Gaussian beam sizes $\theta_{\rm b}$ are listed in Table\,1.
Results are presented for 3 values of the density profile index $\xi$: 
an SIS model with $\xi = -2$, and two more realistic models with 
$\xi = -1.6$ and $-1.4$, which straddle the value $\xi = -1.5$ derived 
from N-body simulations by Moore et al.\ (1998). 
For the more realistic models the effective bias values are less than 
for the SIS case, but in most cases  
positive magnification bias is expected, at least in the innermost demagnified 
regions. 
Some of the bias values listed in Table\,1 are less than unity, corresponding 
to a reduction in the surface density of images. This reduction is greatest 
for the 
largest beamsize, where the positive bias in the central demagnified region 
is counteracted by the negative bias in the surrounding region where 
$\mu > 1$, and for both less centrally concentrated clusters and steeper 
source counts. Within the innermost regions of all the cluster images, 
positive 
bias would still be expected in all cases. 

It is possible, but not currently certain, that the ultradeep 
submillimetre-wave counts could have different slopes at different 
wavelengths. Hence, a differential magnification bias could be detected as a 
function of colour, a submillimetre-wave counterpart to the depletion 
signal detected in optical--near-infrared observations by Gray et al.\ (2000). 

\begin{table} 
\caption{Values of the magnification bias parameter 
$\bar B$ expected in the 
innermost region of a rich cluster with an 
an Einstein radius 
$\theta_{\rm E} = 40$\,arcsec, for three different values of the power-law 
indices of the background galaxy count $\alpha$, the inner radial density 
profile $\xi$, and the FWHM diameter of Gaussian primary beams 
$\theta_{\rm b}$. Values of 
$\theta_{\rm b} = 36$, 18 and 9\,arcsec 
correspond to the beams of the SMA at 850\,$\umu$m, and of ALMA at 850 and 
450\,$\umu$m respectively. The results in a simple approximation to the SIS 
magnification distribution, as marked by $\xi \simeq -2$, with   
$\bar B \propto (\theta_{\rm E}/\theta_{\rm b})^{2+\alpha}$ 
are also listed.
}
{\vskip 0.75mm}
{$$\vbox{
\halign {\hfil #\hfil && \quad \hfil #\hfil \cr
\noalign{\hrule \medskip}
$\theta_{\rm b}$ /& $\alpha$ & $\xi$ & $\bar B$ & 
$\theta_{\rm b}$ /& $\alpha$ & $\xi$ & $\bar B$ \cr
arcsec & & & & arcsec & & & \cr
\noalign{\smallskip \hrule \smallskip}
36 & $-1.84$ & $-2.0$ & 1.04 & 36 & $-1.84$ & $-1.6$ & 0.76 \cr
36 & $-1.84$ & $-1.4$ & 0.64 & 36 & $-1.84$ & $\simeq -2$ & 1.16 \cr
36 & $-1.52$ & $-2.0$ & 1.24 & 36 & $-1.52$ & $-1.6$ & 0.49 \cr
36 & $-1.52$ & $-1.4$ & 0.30 & 36 & $-1.52$ & $\simeq -2$ & 1.63 \cr
36 & $-1.30$ & $-2.0$ & 1.53 & 36 & $-1.30$ & $-1.6$ & 0.42 \cr
36 & $-1.30$ & $-1.4$ & 0.20 & 36 & $-1.30$ & $\simeq -2$ & 2.13 \cr
18 & $-1.84$ & $-2.0$ & 1.24 & 18 & $-1.84$ & $-1.6$ & 0.85 \cr
18 & $-1.84$ & $-1.4$ & 0.73 & 18 & $-1.84$ & $\simeq -2$ & 1.30 \cr
18 & $-1.52$ & $-2.0$ & 2.03 & 18 & $-1.52$ & $-1.6$ & 0.75 \cr
18 & $-1.52$ & $-1.4$ & 0.47 & 18 & $-1.52$ & $\simeq -2$ & 2.27 \cr
18 & $-1.30$ & $-2.0$ & 3.00 & 18 & $-1.30$ & $-1.6$ & 0.85 \cr
18 & $-1.30$ & $-1.4$ & 0.42 & 18 & $-1.30$ & $\simeq -2$ & 3.46 \cr
9.0 & $-1.84$ & $-2.0$ & 1.42 & 9.0 & $-1.84$ & $-1.6$ & 1.06 \cr
9.0 & $-1.84$ & $-1.4$ & 0.93 & 9.0 & $-1.84$ & $\simeq -2$ & 1.45 \cr
9.0 & $-1.52$ & $-2.0$ & 3.01 & 9.0 & $-1.52$ & $-1.6$ & 1.48 \cr
9.0 & $-1.52$ & $-1.4$ & 0.92 & 9.0 & $-1.52$ & $\simeq -2$ & 3.17 \cr
9.0 & $-1.30$ & $-2.0$ & 5.25 & 9.0 & $-1.30$ & $-1.6$ & 2.18 \cr
9.0 & $-1.30$ & $-1.4$ & 1.01 & 9.0 & $-1.30$ & $\simeq -2$ & 5.62 \cr
\noalign{\smallskip \hrule}
\noalign{\smallskip}\cr}}$$}
\end{table}  

Prior to ALMA being commissioned, it will be interesting to search for this 
effect using the SMA. At 345-GHz the best resolution of the SMA is expected 
to be 0.25\,arcsec, and a 1-$\sigma$ sensitivity of 1\,mJy is expected  
in an 8-hour integration. The resolution is thus probably too coarse, and 
the sensitivity insufficiently great to exploit the demagnification bias 
effect to the full.  

Although the magnification bias can increase the surface density of 
detectable galaxies in the innermost parts of clusters, this increase 
corresponds to a reduction in the fraction of the background radiation 
intensity that is resolved in detected galaxies. In order to detect the greatest 
possible proportion of the submillimetre-wave background radiation 
intensity, ALMA observations of the most strongly magnified regions of 
clusters of galaxies are still required.  

\subsection{Potential caveats} 

`Demagnification bias' could make observations of extremely faint counts 
of galaxies significantly easier using ALMA, if the slope of the faint 
counts is shallow, $\alpha \gs -1.5$, and the density profile of the cluster 
is centrally 
peaked, $\xi \ls -1.5$. However, it can only be 
exploited if both the angular resolution of the resulting images is  
sufficient to allow adjacent lensed images to be resolved, and the confusion 
limit is sufficiently deep. The physical extent of the lensed
background galaxies must also be small enough to avoid them overlapping 
on the sky, and 
there must be no strong emission or absorption from the lensing cluster  
to mask the demagnified sources. 

\subsubsection{Resolution, confusion and source size} 

Several tens of resolution elements per source within the primary beam are 
required to satisfy both the confusion and resolution requirements. This 
relates to a resolution of order 0.1\,arcsec, which will easily be achieved 
using ALMA at a wavelength of 850\,$\umu$m on even a relatively short 2-km 
baseline. At the maximum planned 10-km baseline, the resolution at 850 and 
450\,$\umu$m is considerable better, 0.02 and 0.01\,arcsec respectively. The 
source sizes should also be sufficiently small. There is evidence for large 
halos of cold gas around the most luminous high-redshift dust-enshrouded 
galaxies (Papadopoulos et al.\ 2001), but other sources are known to be 
smaller than a few arcseconds in size (Frayer et al.\ 1998, 1999, 2000;
Downes et al.\ 1999; Lutz et al.\ 2001). They will also
be reduced in extent by demagnification. 

\subsubsection{Contamination from cluster emission} 

A great advantage of the $K$-correction in the submillimetre waveband is 
that high-redshift galaxies are as easy to detect as their low-redshift 
counterparts (Blain \& Longair 1993). 
This is verified by the lack of a significant fraction of 
low-redshift galaxies detected in SCUBA surveys (Smail et al.\ 2001), with 
the exception of 2 cD galaxies in the centres of the target clusters 
containing powerful cooling flows (Edge et al.\ 1999). 

Submillimetre-wave emission from the interstellar medium (ISM) in cD 
galaxies is intrinsically interesting, as it could reveal the fate of gas that 
cools from the X-ray emitting intracluster medium. However, it could also 
mask the demagnified images of background galaxies in an extremely deep 
ALMA observation. Even magnified radial-arc images can be masked by 
starlight from cD galaxies in optical {\it HST} images (Smith et al.\ 2001). 

It is likely that any contaminating emission from the cD galaxy ISM could be 
subtracted reliably from the ALMA images. ALMA has the sensitivity to 
resolve this emission in several different CO transitions. The cD emission is 
also likely to be spread over an area of at least several square arcseconds, 
and so should have a reduced surface brightness as compared with the 
images of background galaxies. In addition, the continuum radiation from 
the background galaxies will undergo molecular line absorption at the 
redshift of the cD galaxy, and so by searching for these narrow 
absorption lines, it should be possible to further discriminate between 
emission from background images and the cD galaxy. 
Absorption by gas 
within the cluster and cD galaxy is not likely to be important away from 
the frequencies of these discrete absorption lines.  

\subsection{Determining the central cluster potential and the  
geometry of the Universe} 

Based on the observed positions of many sets of multiple images of 
background galaxies detected using ALMA, a significant fraction of which 
will be certain to be identified correctly, with redshifts determined 
serendipitously from the detection of CO lines (Blain et al.\ 2000; Blain
2001), it should be possible to reconstruct accurately the gravitational 
potential very close to the cluster core, and so reveal the density profile of 
both visible and dark matter. 
This is impossible using optical observations, as starlight from the cD 
galaxy masks the lensed images. The detection of any magnified radial-arc 
images within 10--20\,arcsec of the cluster core, see Fig.\,2, 
would also provide  
useful constraints on the gradient of the local potential.

The magnified counter-images to the demagnified images detected in the 
innermost regions of the cluster are expected to lie close to critical lines. 
Knowledge of their positions, especially of those sets of multiple images 
confirmed using CO redshifts, can be used to construct exact mass models 
of the inner few arcminutes of the lensing cluster. In addition, several sets 
of multiple images with redshifts could be used to investigate the 
geometry of the Universe by finding the relative geometrical distances 
between the observer, cluster and source; compare with the triplet method of 
Gautret, Fort \& Mellier (2000) for weak lensing.  

The time required to complete such a multiple imaging survey should be 
comparable to the time required to image the demagnified central region 
of the cluster. The magnified counter-images are expected to lie close to 
critical lines, and so could be detected in a series of about 20 ALMA images 
forming a ring around the centre of a cluster. These images would be 
significantly brighter than the central demagnified images, and so a shorter 
integration time per field 
would be required. Over many years, it would be desirable 
to build up multi-wavelength ALMA images of the entire central regions of 
clusters; however, maps of both the innermost core and the critical lines, 
generating a bullseye image, with a central ultradeep field in the core, 
surrounded by 
an annulus of shallower observations tracing the critical lines several tens 
of arcseconds away, are 
the most urgent requirements.  

This type of survey will not provide a fully representative sample of the 
distant Universe, as it is necessarily limited to 1-arcmin-diameter pencil 
beams passing through the cores of at most several hundred rich 
clusters at intermediate redshifts. However, it will provide an increased 
efficiency for the determination of the very faintest submillimetre-wave 
counts. 

\section{Conclusions} 

At the depths suitable for the detection of a number of galaxies within the 
primary beam of ALMA, the differential submillimetre-wave source counts, 
$N \propto S^\alpha$ are likely to 
be rather flat, with $\alpha \gs -1.8$. In the significantly demagnified
regions within about 10\,arcsec of the cores of rich clusters of galaxies, 
this corresponds to an increase in the surface density of faint 
sub-100\,$\umu$Jy galaxies. Ultradeep ALMA images of the innermost 
regions of cluster cores could thus speed the detection of 
the population of normal, $L^*$ high-redshift 
galaxies, if clusters have a central density profile index $\xi \ls -1.5$ 
and the faint count slope $\alpha \gs -1.5$. 
An ultradeep pencil-beam redshift survey would be provided as a 
by-product, from the simultaneous detection of CO emission and absorption 
lines in the spectra of the detected galaxies. By detecting several sets of 
magnified counter-images to these sources, which would lie at radii of 
order 1\,arcmin from the cluster core, it should be possible to provide 
accurate measures of both the central cluster potential and the 
geometry of the Universe. 

\section*{Acknowledgements} 

Thanks to Ole M\"oller, Priya Natarajan, Kate Quirk and an anonymous 
referee for valuable comments 
on the manuscript. The author acknowledges generous support from 
the Raymond and Beverly Sackler Foundation  
as part of the Deep Sky Initiative programme at the IoA.

\end{document}